%% file: main.tex
\documentclass{article}

\usepackage{hyperref}

\usepackage{booktabs}
\usepackage{multirow}
\usepackage{makecell}

\usepackage{pgfplots}
\pgfplotsset{compat=1.9}
\usetikzlibrary{patterns}
\usepgfplotslibrary{patchplots}
\usepgfplotslibrary{groupplots}

\usepackage{caption}
\usepackage{subcaption}

\usepackage[preprint]{icml2025}

\usepackage{amsmath}
\usepackage{amssymb}
\usepackage{mathtools}
\usepackage{amsthm}

\usepackage[capitalize,noabbrev]{cleveref}

\theoremstyle{plain}

\theoremstyle{definition}

\theoremstyle{remark}

\newcommand{\squishlist}{
	\begin{list}{$\bullet$}
		{ \setlength{\itemsep}{1pt}
			\setlength{\parsep}{1pt}
			\setlength{\topsep}{2.5pt}
			\setlength{\partopsep}{0.5pt}
			\setlength{\leftmargin}{1em}
			\setlength{\labelwidth}{1em}
			\setlength{\labelsep}{0.6em}
		}
	}
	\newcommand{\squishend}{
	\end{list}
}

\newcommand{\name}{PilotANN}

\begin{document}

\twocolumn[
    \icmltitle{\name{}: Memory-Bounded GPU Acceleration for Vector Search}




    \begin{icmlauthorlist}
        \icmlauthor{Yuntao Gui}{cuhk}
        \icmlauthor{Peiqi Yin}{cuhk}
        \icmlauthor{Xiao Yan}{cpii}
        \icmlauthor{Chaorui Zhang}{huawei}
        \icmlauthor{Weixi Zhang}{huawei}
        \icmlauthor{James Cheng}{cuhk}
    \end{icmlauthorlist}

    \icmlaffiliation{cuhk}{The Chinese University of Hong Kong, Hong Kong SAR}
    \icmlaffiliation{cpii}{Centre for Perceptual and Interactive Intelligence, Hong Kong SAR}
    \icmlaffiliation{huawei}{Theory Lab, 2012 Labs of Huawei Technologies Co. Ltd.}
    \icmlcorrespondingauthor{Yuntao Gui}{ytgui@cse.cuhk.edu.hk}

    \icmlkeywords{Machine Learning, Approximate Nearest Neighbor Search, GPU Programming, High Performance Compute}

    \vskip 0.3in
]



\printAffiliationsAndNotice{}  

\begin{abstract}
    Approximate Nearest Neighbor Search (ANNS) has become fundamental to modern deep learning applications, having gained particular prominence through its integration into recent generative models that work with increasingly complex datasets and higher vector dimensions.
    Existing CPU-only solutions, even the most efficient graph-based ones, struggle to meet these growing computational demands, while GPU-only solutions face memory constraints.
    As a solution, we propose \name{}, a hybrid CPU-GPU system for graph-based ANNS that utilizes both CPU's abundant RAM and GPU's parallel processing capabilities.
    Our approach decomposes the graph traversal process of top-$k$ search into three stages: GPU-accelerated subgraph traversal using SVD-reduced vectors, CPU refinement and precise search using complete vectors.
    Furthermore, we introduce fast entry selection to improve search starting points while maximizing GPU utilization.
    Experimental results demonstrate that \name{} achieves $3.9 - 5.4 \times$ speedup in throughput on 100-million scale datasets, and is able to handle datasets up to $12 \times$ larger than the GPU memory.
    We offer a complete open-source implementation of \name{}: \url{https://github.com/ytgui/PilotANN}.
\end{abstract}

\input{section/1-introduction.tex}

\input{section/2-background.tex}

\input{section/3-overview.tex}

\input{section/4-piloting.tex}

\input{section/5-routing.tex}

\input{section/6-evaluation.tex}

\input{section/7-related.tex}

\input{section/8-conclusion.tex}

\input{section/9-impact.tex}

\bibliography{main}
\bibliographystyle{icml2025}

\clearpage
\appendix
\input{section/10-appendix.tex}


\end{document}

%% file: section/1-introduction.tex
\section{Introduction}\label{sec:introduction}

Approximate Nearest Neighbor Search (ANNS) is a fundamental vector search technique that efficiently identifies similar items in high-dimensional vector spaces. Given a query vector, ANNS returns its $k$-nearest neighbors by making controlled accuracy and search speed trade-offs.
This balance between accuracy and speed has made ANNS the de facto solution for large-scale similarity search problems where exact nearest neighbor computation would be computationally prohibitive.
Traditionally, ANNS has served as the backbone for retrieval engines~\cite{kulis2009search, gordo2016search} and recommendation systems~\cite{covington2016youtube, he2017cf, yang2020twotower}.
More recently, ANNS has gained renewed prominence through its integration into generative AI systems, where it enhances both truthfulness and efficiency, e.g., retrieval augmented generation~\cite{lewis2020rag, blattmann2022diffusion}, and semantic cache~\cite{bang2023gptcache}.

\input{resource/1-motivation.tex}

The computational requirements of ANNS continue to grow with recent advances in generative AI, particularly as modern Transformer-based architectures work with higher-dimensional embeddings and larger-scale datasets~\cite{devlin2018bert, dosovitskiy2021vit, radford2021clip}.
While deep learning systems can be easily replicated and scaled horizontally due to their stateless nature, ANNS typically remains centralized, making single-machine throughput a critical bottleneck.
Using 100-million scale datasets with dimensions ranging from 96 to 768, we observe that state-of-the-art CPU implementation of HNSW~\cite{malkov2018hnsw} struggles to maintain high query throughput as vector dimensions increase (Figure~\ref{fig:motivation}).
Specifically, for the LAION dataset with 768-dimensional vectors, CPU-only HNSW can only process 2.1K queries per second (QPS).
This low throughput becomes particularly problematic for production services that need to handle thousands of concurrent requests, where maintaining high processing throughput is essential.

Previous attempts to enhance graph-based ANNS efficiency have explored several approaches.
One direction focuses on developing fine-grained graph construction methods~\cite{fu2019nsg, lu2021hvs}.
While these methods improve search efficiency, they often require intensive computational resources during the build phase and face scalability challenges with large-scale datasets~\cite{chen2021spann}.
Another approach employs quantization techniques~\cite{jegou2010ivfadc, zhou2012scalar, douze2018linkandcode} to compress vector representations, reducing memory footprint and computational costs.
However, quantization inevitably introduce accuracy degradation, presenting a fundamental trade-off on search accuracy.

GPU acceleration represents a promising direction for enhancing ANNS performance~\cite{zhao2020song, zhu2022rtnn, ootomo2023cagra}.
However, leveraging GPUs is a non-trivial endeavor despite their massive parallel processing capabilities.
The primary limitation is GPU memory capacity, which constrains the size of datasets that can be processed.
NVIDIA's state-of-the-art GPU-based ANNS library CAGRA~\cite{ootomo2023cagra}, requires expensive GPU hardware with large memory capacity (e.g., A100) to handle moderate-sized datasets.
While unified virtual memory (UVM) has been explored as a solution for memory constraints, previous work has shown its ineffectiveness in non-graph ANNS~\cite{zhang2024rummy}.
Graph-based ANNS presents additional challenges due to its dynamic nature and irregular memory access patterns (see \S\ref{subsec:bg-traversal}), making efficient GPU utilization particularly difficult.

These limitations have created a clear need for a hybrid approach that can utilize both CPU and GPU capabilities effectively. In response, we present \name{}, a hybrid system for graph-based ANNS designed to harness both the parallel processing power of the GPU and the abundant RAM available on the CPU. Our approach specifically addresses two fundamental challenges:
\squishlist
\item \textbf{C1: GPU memory boundaries.}
The limited memory capacity of a GPU, typically in the tens of gigabytes, is significantly smaller than that of the CPU, which can exceed hundreds of gigabytes, thus restricting the applicability of ANNS on the GPU.
\item \textbf{C2: Limited computational density.}
GPUs are optimized for complex matrix-matrix operations (e.g., GEMM), but ANNS primarily relies on simpler pairwise distance calculations, resulting in low computational density on the GPU.
\squishend

\name{} tackles these challenges through two key innovations.
First, we introduce \textbf{Multi-stage ANNS Processing} (\S\ref{sec:piloting}), which decomposes the computationally intensive top-$k$ search problem into complementary components: leveraging GPU power to efficiently identify candidates on smaller sub-graphs, followed by refinement and precise traversal on the CPU, where the complexity of ANNS is significantly reduced.
Second, we develop \textbf{Fast Entry Selection} (\S\ref{sec:routing}), a novel method that provides high-quality entry points for search, resulting in increased computational density on the GPU.
Our experimental evaluation, conducted using only one NVIDIA A10 GPU (24 GB), demonstrates significant performance improvements.
We achieve a throughput speedup of $3.9-5.4 \times$ for top-10 searches, while handling datasets and graph index up to $12 \times$ larger than the GPU's memory capacity.

The contributions of this work are threefolds:
\squishlist
\item We introduce a novel hybrid architecture that effectively enables GPU acceleration for CPU-based ANNS.
\item We develop fast entry selection, a GPU-efficient method for optimizing entry point selection.
\item We provide comprehensive empirical evidence of \name{}'s effectiveness on real-world datasets.
\squishend

%% file: resource/1-motivation.tex
\begin{figure}[t]
    \centering
    \scriptsize
    \begin{tikzpicture}
        \begin{axis}[
                ybar,
                ymode=log,
                width=0.40\textwidth,
                height=0.21\textwidth,
                enlarge x limits=0.20,
                ylabel={QPS (K)},
                nodes near coords,
                point meta=explicit symbolic,
                nodes near coords style={above},
                symbolic x coords={DEEP, T2I, WIKI, LAION},
                xtick=data, bar width=8pt, ymin=1.0, ymax=250.0,
                legend style={
                        legend columns=2, legend cell align={left}
                    },
                legend image code/.code={
                        \draw[#1] (0cm,-0.1cm) rectangle (0.2cm,0.1cm);
                    }
            ]
            \addplot+ [
                color=black, fill=orange!60
            ] table[x=x, y=y, row sep=\\] {
                    x           y               \\
                    DEEP        11.2            \\
                    T2I         1.3             \\
                    WIKI        4.2             \\
                    LAION       2.1             \\
                };
            \addplot+ [
                color=black, fill=cyan!60,
            ] table[x=x, y=y, meta=meta, row sep=\\] {
                    x           y          meta      \\
                    DEEP        43.7       3.9x      \\
                    T2I         6.6        5.1x      \\
                    WIKI        17.8       4.2x      \\
                    LAION       11.3       5.4x      \\
                };
            \legend{HNSW, \name{}}
        \end{axis}
    \end{tikzpicture}
    \caption{Recall@10=0.90 QPS on 100 million datasets.}
    \label{fig:motivation}
\end{figure}
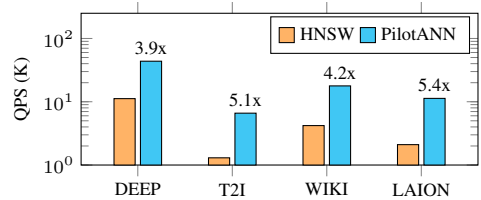

%% file: section/2-background.tex
\section{Background}\label{sec:background}

In this part, we introduce the basics of ANNS and graph traversal algorithm to facilitate the subsequent discussions.

\subsection{ANNS Methods}\label{subsec:bg-anns}

The goal of ANNS is to efficiently find the data points in a given dataset $X$ that are closest to a query point $q$ according to a distance metric~\cite{chen2021spann, peng2023speedann, douze2024faiss}.
Modern ANNS methods can be broadly categorized into two main approaches: non-graph methods based on coarse clustering, and graph-based methods that rely on fine-grained graph connectivity.

\textit{Traditional non-graph methods} employ coarse clustering as their foundational principle, partitioning the search space to reduce computational complexity during query time.
This includes techniques like space-partitioning KD-tree~\cite{anan2008kdtree} that recursively divide the space into hierarchical regions, and locality-sensitive hashing~\cite{gionis1999lsh} that groups similar vectors into buckets.
The most direct application of clustering is the Inverted File (IVF) method~\cite{jegou2011ivfadcr}, which explicitly partitions vectors using k-means into inverted lists.
Advanced variants like IMI~\cite{babenko2015imi} employ product clustering to create finer partitions, while IVFADC~\cite{schutze2008introduction, jegou2010ivfadc} combines clustering with residual quantization.
However, these coarse clustering methods face an inherent trade-off between search complexity and partition granularity.

\textit{Graph-based methods} take a fundamentally different approach by constructing a navigable graph structure~\cite{hajebi2011knngraph, malkov2014nsw}, where each data point is represented as a node and edges connect neighboring nodes. During search, these methods perform graph traversal to quickly locate nearest neighbors.
Modern approaches like HNSW~\cite{malkov2018hnsw} organize the proximity graph into multiple layers for coarse-to-fine search, while NSG~\cite{fu2019nsg} optimizes graph connectivity for better search routes.
Studies have consistently shown that graph-based methods achieve state-of-the-art performance in terms of the accuracy-throughput trade-off, requiring fewer distance computations than traditional coarse clustering methods to achieve the same recall rates~\cite{douze2018linkandcode, chen2021spann}.
This superior performance stems from their ability to capture fine-grained neighborhood connections and perform guided graph traversal, which makes graph-based approaches highly optimized.
Exploring further improvements on graph-based methods remains an active area of research~\cite{zhao2020song, ootomo2023cagra, karthik2024bang}.

\subsection{Revisiting Graph Traversal}\label{subsec:bg-traversal}

\input{resource/2-algorithm.tex}

Graph-based ANNS typically employs a heuristic greedy search algorithm~\ref{alg:greedy-search}.
This algorithm explores the graph efficiently by maintaining a restricted number of the most promising candidate solutions at each step, thereby limiting computational complexity.

The algorithm begins by constructing a candidate list $C$ (line 3) containing the most promising solutions identified during the search process.
This list comprises $ef$ candidates, from which the top-$k$ results ($k \le ef$) can be obtained later. Here the $ef$ serves as a parameter to control search accuracy and speed.
In each subsequent iteration, the search begins by expanding the best unchecked candidate in $C$ (line 5-6), computing distances of unvisited neighboring nodes to the query vector $q$ (line 8), and updating $C$ as potential solutions (line 9).
This iterative process continues until all nodes in $C$ have been checked (line 4), indicating that no further improvement is possible within the current search space.

The greedy search algorithm exhibits two key characteristics.
First, at each neighbor expansion iteration, the algorithm selects the best candidate from the dynamically changing $C$, making the traversal path unpredictable. This dynamic nature with \textbf{C1} motivates most of the system design.
Second, The algorithm primarily performs vector-to-vector distance calculations (line 8), which are simple element-wise operations that cannot utilize the GPU's matrix multiplication (GEMM) capabilities, i.e., \textbf{C2}.

%% file: resource/2-algorithm.tex
\begin{algorithm}[t]
    \caption{ANNS by Greedy Search}
    \begin{algorithmic}[1]
        \STATE {\bfseries Input:} graph $G(V,E)$, query $q$, entry points $EP$, candidate size $ef$
        \STATE {\bfseries Output:} $C$ contains $ef$ nearest neighbors
        \STATE priority queue $C \gets EP$
        \REPEAT
        \STATE $u \gets$ the first unchecked node in $C$
        \FOR{unvisited $v \in \{v | (u, v) \in E\}$}
        \STATE mark $v$ as visited
        \STATE $d \gets euclidean(q, v)$
        \STATE $C.insert(v)$ w.r.t. $d$
        \ENDFOR
        \STATE $C.resize(ef)$
        \UNTIL{$C$ has no unchecked node}
    \end{algorithmic}
    \label{alg:greedy-search}
\end{algorithm}

%% file: section/3-overview.tex
\section{Overview}\label{sec:overview}

\begin{figure}[t]
    \centering
    \includegraphics[width=0.47\textwidth]{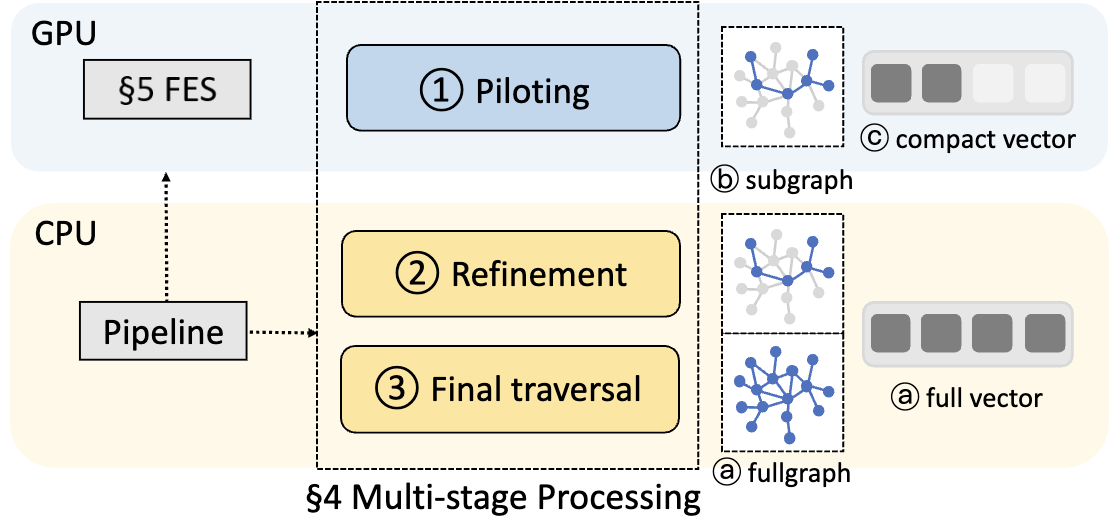}
    \caption{Architecture of \name{}.}
    \label{fig:overview}
\end{figure}

\name{} employs a unique hybrid CPU-GPU processing paradigm that leverages the distinct advantages of both hardwares. Figure~\ref{fig:overview} illustrates the system's architecture and core components.

The key insight of \name{} is to decompose the ANNS process into complementary stages that run on both GPU and CPU.
During preprocessing, given the full index~\textcircled{a}, we sample a smaller subgraph~\textcircled{b} and generate compact vector representations~\textcircled{c}.
The smaller subgraph and compact vectors can fit within GPU memory, while the original full index kept in CPU RAM.
At runtime, \name{} processes queries through our multi-stage processing method (\S\ref{sec:piloting}), where the GPU stage \textcircled{1} enables efficient initial exploration, and the CPU stages \textcircled{2}\textcircled{3} benefit from reduced search complexity.
Moreover, fast entry selection (FES \S\ref{sec:routing}) is designed to improve the quality and speed of initial starting point selection.
Together, \name{} is able to achieve high throughput while effectively processing datasets that significantly exceed GPU memory capacity.

%% file: section/4-piloting.tex
\section{GPU Piloting: A Hybrid Approach}\label{sec:piloting}

\subsection{Multi-stage Processing of ANNS}\label{subsec:multi-stage}

\name{} introduces a novel approach to vector search through a \textbf{``staged data ready processing''} paradigm.
Unlike traditional GPU-enabled systems that adhere to a ``move data for computation'' model, our method minimizes data movement by ensuring data readiness across processing stages.
This design choice is particularly crucial due to the inherent characteristics of graph traversal algorithm shown in \S\ref{subsec:bg-traversal}.
During iterative neighbor expansions, the search path becomes data-dependent and unpredictable, which would necessitate intensive data transfers over PCIe in conventional architectures.

Our method breaks down the search into three distinct stages: (i) GPU piloting with subgraph and reduced vectors, (ii) residual refinement with subgraph and full vectors, and (iii) final traversal with full graph and vectors.
Beginning with a query and entry points, the system sequentially executes these stages to locate top-$k$ matching results.

\textbf{\textcircled{1} GPU piloting.}
In the first stage, we employ dimensionality reduction and graph sampling to enable GPU-based traversal under memory constraints.
For dimensionality reduction, we apply singular value decomposition (SVD) to the original vectors: $X = U \Sigma V^T$, where $V^T V = I$ ensures the orthogonality of the transformation, preserving Euclidean distances.
Each vector is decomposed into two components: $\hat{x} = \{ x_{primary}, x_{residual} \}$, where $x_{primary} = \{ U_{1,\dots,d} \Sigma_{1,\dots,d}$ \}, captures the principal components with $d$ highest singular values, and $x_{residual}$ contains the remaining components.
The graph sampling process employs uniform node-wise sampling to select seed nodes, followed by 1-hop neighbor expansion to include frontier nodes, until reaching a target sampling ratio; the sampled nodes are then reconnected using the same graph construction algorithm as the original index.

\textbf{\textcircled{2} Residual refinement.}
This intermediate stage enhances the GPU results by incorporating the previously omitted residual vector components.
For each candidate identified in \textcircled{1}, we calculate complete distance by combining the GPU-computed $x_{primary}$ distance with the $x_{residual}$ distance.
The stage then performs a limited graph traversal (2 iterations) on the subgraph, re-ranking candidates based on their full-dimensional distances and generating a visited table.
This refinement process produces two key outputs: a more accurately ranked candidates and a visitation history that guides the subsequent complete graph exploration.

\textbf{\textcircled{3} Final traversal.}
The final stage performs a complete graph search using full-dimensional vectors, ensuring search quality and completeness.
Building upon the previous stages' outputs, it proceeds a traditional greedy search algorithm \S\ref{subsec:bg-traversal}  with two key advantages: a pre-populated visited table and high-quality initial candidates.
By reusing the visited table, the search avoids revisiting previously explored nodes.
The quality of starting points obtained from earlier stages significantly improves traversal efficiency, see~\S\ref{subsec:stage-analysis}.

\textbf{Summary.}
Our design is cost-effective as it requires only a single GPU, while scaling well across vector dimensions and graph complexity.
Data transfer overhead is minimal -- only the query vector moves to GPU memory initially, and a small candidate set (typically less than 1KB per query) returns to CPU after \textcircled{1}.
The design ensures search quality through graceful degradation: if only the final stage operates, the system functions as a traditional greedy search, while bringing speedups when all stages work together.

\subsection{Understanding the Performance}\label{subsec:stage-analysis}

\input{resource/4-analysis.tex}

Our approach demonstrates effectiveness through two aspects.
First, the GPU stage \textcircled{1} can significantly reduces search complexity for subsequent CPU stages \textcircled{2}\textcircled{3}.
Second, the GPU's massive parallel processing capabilities are essential for our approach, as they enable efficient graph traversal that ensures multi-stage overheads do not compromise its benefits.
We validate these claims through the following performance analysis.

\textbf{Reduced search complexity.}
Our method builds upon a simple but powerful observation: graph-based top-$k$ search becomes much more efficient when it begins with some ground truth results already known.
To measure this effect, we conducted experiments with HNSW using 32 neighbors per node, comparing searches initialized with or without partial ground truth.
Specifically, for the former, we construct initial search candidate of size $ef$ by combining $\tau$ known correct results with $ef - \tau$ randomly selected nodes.

Figure~\ref{fig:analysis-tau} demonstrates the substantial benefits of this approach.
When starting with $\frac{\tau}{ef} = \frac{1}{4}, \frac{1}{8}$, the search only requires 39.9\% and 48.1\% distance calculations to reach a recall of 0.90 compared to search without ground truth.
This improvement happens because the search immediately focuses on promising areas of the graph, avoiding wasted effort in exploring less relevant regions. We observe similar improvements across different datasets, showing its generalizability.
In the proposed design, the first two stages \textcircled{1}\textcircled{2} utilizing GPU's parallel processing power to rapidly identify high-quality neighbors that serve as approximate ground truth, accelerating the final stage \textcircled{3}.

This benefit only materialize when identifying a sufficient proportion of ground truth neighbors.
We define \textit{acceleration threshold} as the minimum ratio $\frac{\tau}{ef}$ needed to achieve meaningful speedup (e.g., $2\times$).
As Figure~\ref{fig:analysis-threshold} reveals, this threshold varies across datasets, typically requiring 15-21\% of ground truth results.
This minimum threshold requirement guides our system design -- we employ both graph sampling and vector dimensionality reduction to maximize the subgraph size that can fit within limited GPU memory.

\begin{table}[t]
    \small
    \centering
    \setlength{\tabcolsep}{4.0mm}
    \caption{Stage breakdown on LAION-1M.}
    \begin{tabular}{c|cccc}
        \toprule
        \textbf{Method} & \textbf{Stage \textcircled{1}} & \textbf{Stage \textcircled{2}} & \textbf{Stage \textcircled{3}} \\
        \midrule
        HNSW            & -                              & -                              & 668.8                          \\
        Multi-stage     & 574.2                          & 44.2                           & 189.0                          \\
        \bottomrule
    \end{tabular}
    \label{tab:decompose}
\end{table}

\textbf{Stage breakdown.}
A breakdown of distance calculations across stages reveals why GPU acceleration is fundamental to our design (Table~\ref{tab:decompose}).
While the baseline HNSW requires 668.8 calculations to reach a recall of 0.9, our method distributes its workload across three stages: 574.2 calculations in the initial GPU stage, followed by 44.2 and 189.0 calculations in two CPU stages.
The CPU-executed portions, totaling 233.2 calculations on \textcircled{2}\textcircled{3}, are $3.3 \times$ smaller than the baseline, which is the key to our acceleration.
Although the Multi-stage involves more calculations in total, the GPU's parallel processing capabilities handle the increased workloads of \textcircled{1} efficiently.
Specifically, we measure vector-to-vector distance computation on CPU and GPU, the GPU is capable of handling $82 \times$ more computations over a single CPU core.
We further optimize system throughput by processing queries in batches and pipelining their execution across CPU and GPU stages.

\subsection{Methodology Details}\label{subsec:stage-details}

\textbf{Subgraph management.}
The subgraph representation employs a modified Compressed Sparse Row (CSR) format that optimizes memory usage  while maintaining efficient mapping between the subgraph and full graph.
Rather than completely removing nodes excluded during sampling, we retain their presence in the CSR structure but remove their connectivity information and embedding vectors. These nodes are represented with zero out-degree, and any incoming edges are pruned.
This design makes the data structure on the GPU have a little redundancy (these zero-degree nodes), however, it avoids the computational overhead of node ID mapping between subgraph and fullgraph representations, resulting in improved CPU performance.

\textbf{Visited table management.}
Our GPU kernel utilizes bloom filters to track visited nodes, with the filter states stored in shared memory. This design eliminates the need for dynamic memory allocation and reduces DRAM access overhead.
While bloom filters are known to produce false positives -- incorrectly marking unvisited nodes as visited -- this limitation does not compromise our system's correctness.
Our multi-stage processing pipeline ensures accuracy by refining GPU results on the CPU side, where any false positive nodes are properly revisited.

%% file: resource/4-analysis.tex
\begin{figure}[t]
    \centering
    \scriptsize
    \begin{minipage}[b]{0.23\textwidth}
        \begin{tikzpicture}
            \begin{axis}[
                    xlabel=Recall@10,
                    ylabel=Calculation (K),
                    xmin=0.75, xmax=1.00,
                    ymin=0.05, ymax=1.75,
                    scale only axis=true,
                    width=0.75\textwidth,
                    height=0.35\textwidth,
                    legend style={
                            anchor=south west, at={(-0.01, 1.05)}, legend columns=3
                        }
                ]
                \addplot+ [color=orange!90, mark=x] coordinates {
                        (0.767, 0.397) (0.873, 0.593) (0.925, 0.927)
                        (0.944, 1.254) (0.950, 1.558)
                    };
                \addplot+ [color=cyan!90, mark=triangle] coordinates {
                        (0.862, 0.241) (0.921, 0.439) (0.952, 0.800)
                        (0.962, 1.134) (0.964, 1.453)
                    };
                \addplot+ [color=blue!90, mark=diamond] coordinates {
                        (0.883, 0.237) (0.932, 0.436) (0.959, 0.794)
                        (0.964, 1.124) (0.967, 1.441)
                    };
                \legend{0, $\frac{1}{8}$, $\frac{1}{4}$}
            \end{axis}
        \end{tikzpicture}
        \caption{Calculation requirements of LAION-1M under different $\frac{\tau}{ef}$ conditions.}
        \label{fig:analysis-tau}
    \end{minipage}
    \hfill
    \begin{minipage}[b]{0.23\textwidth}
        \begin{tikzpicture}
            \begin{axis}[
                    ybar,
                    width=1.15\textwidth,
                    height=0.80\textwidth,
                    enlarge x limits=0.15,
                    ylabel={Threshold},
                    symbolic x coords={DEEP, T2I, WIKI, LAION},
                    xtick=data, bar width=7pt, ymin=0.00, ymax=0.32,
                    legend style={
                            anchor=north, at={(0.5, 1.40)},
                            legend columns=2, legend cell align={left}
                        },
                    legend image code/.code={
                            \draw[#1] (0cm,-0.1cm) rectangle (0.2cm,0.1cm);
                        }
                ]
                \addplot+ [color=black, fill=orange!60] coordinates {
                        (DEEP, 0.149) (T2I, 0.213) (WIKI, 0.129) (LAION, 0.105)
                    };
            \end{axis}
        \end{tikzpicture}
        \caption{Acceleration thresholds to obtain $2 \times$ distance calculation savings.}
        \label{fig:analysis-threshold}
    \end{minipage}
\end{figure}

%% file: section/5-routing.tex
\section{Fast Entry Selection}\label{sec:routing}

In graph-based ANNS, the search begins with entry points before executing graph traversal.
These entry points can be predefined nodes~\cite{malkov2018hnsw} or randomly selected points~\cite{fu2019nsg}.
We introduce Fast Entry Selection (FES), a novel method optimized for GPU execution that operates before \S\ref{sec:piloting}.
FES employs GEMM-like high-density distance computations to improve entry point quality without compromising search speed, thereby enhancing overall system efficiency.

\begin{table}[t]
    \small
    \centering
    \setlength{\tabcolsep}{2.5mm}
    \caption{Complexity of distance computations.}
    \begin{tabular}{c|ccc}
        \toprule
        \textbf{Method} & \textbf{Comp.}    & \textbf{Mem. read}  & \textbf{Density}        \\
        \midrule
        Brute force     & $m n d$           & $m d + n d$         & $\frac{m n}{m + n}$     \\
        Graph traversal & $m n d$           & $m d + m n d$       & $\frac{n}{1 + n}$       \\
        \midrule
        FES (general)   & $\frac{m n d}{r}$ & $m d + n d$         & $\frac{m n}{r (m + n)}$ \\
        FES (1 block)   & $m n d$           & $m d + n d$         & $\frac{m n}{m + n}$     \\
        FES (1 query)   & $\frac{n d}{r}$   & $d + \frac{n d}{r}$ & $\frac{n}{1 + n}$       \\
        \bottomrule
    \end{tabular}
    \label{tab:complexity}
\end{table}

\textbf{Computational density.}
We first examine computational density w.r.t. the roofline performance model~\cite{williams2009roofline}.
Given vector dimension $d$, we have $m$ query vectors $Q \in \mathbb{R} ^ {m \times d}$ and $n$ entry vectors $EV \in \mathbb {R} ^ {n \times d}$.
As listed in Table~\ref{tab:complexity}, a straightforward brute force approach computes the distance between each query and entry vector, resulting in an $m \times n$ matrix $D$, where each element represents the distance between one query vector and one entry vector.
Treating the squared euclidean distance computation $euclidean(x_1, x_2) = (x_1 - x_2) ^ 2$ as a single computation, the entry stage involves $m n d$ computations with $m d + n d$ memory reads, the ratio between computation and memory access is $\frac{m n}{m + n}$, i.e., computational density.

During neighbor expansion (lines 6-10 in Algorithm~\ref{alg:greedy-search}), computing the same distances $D$ requires $m d + m n d$ memory reads for query vectors $Q$ and neighbor vectors $V \in \mathbb{R}^{m \times n \times d}$, where each of the $m$ queries has different $n$ neighbors.
This yields a computational density of $\frac{n}{1 + n}$, significantly lower than $\frac{m n}{m + n}$ as $m$ increases.
A comparable scenario during the graph construction process, namely local-join, was also reported~\cite{dong2011nndescent}.
This observation motivates us to explore novel methods that can maintain high computational density while reducing the effective search space.

\textbf{Clustering-based selection.}
Inspired by IVF, we propose a simplified and bucket-aligned approach optimized for entry point selection.
Our method organizes entry vectors into a small number of coarse clusters ($r$), enabling efficient GEMM-like computations on GPUs.
Unlike IVF, which typically employs thousands of buckets with highly variable sizes, our method deliberately uses a small number of coarse clusters ($r <<$ IVF buckets).
For queries $Q$ and entry vectors $EV$, our approach dynamically routes queries to appropriate clusters based on their proximity to cluster centroids.
When a subset of queries $Q_i$ are assigned to cluster $i$, it activates entry vectors $EV_i$ within that cluster for distance computation.
Entry vectors in other clusters are considered too distant and are excluded from the distance calculation for that particular query.

\input{resource/5-algorithm.tex}

\textbf{Allocation-free and tiled FES.}
The final challenge is to implement FES efficiently.
A straightforward implementation might extract both $Q_i$ and $EV_i$ before performing distance computations. However, this approach can lead to memory allocations and copies, potentially reducing performance.
To overcome this limitation, we have developed an allocation-free and tiled GPU kernel, as shown in Algorithm~\ref{alg:fast-routing}.
The algorithm begins by distributing the computation across multiple GPU blocks (line 3), each handling a subset of queries and entry vectors. Within each block, the algorithm statically declares two shared memory blocks to store partial query and entry vectors (line 4).
Next, each block iterates through the entry vectors and queries in a tiled manner (line 7, 12). This tiling approach is crucial for performance as it maximizes memory reads and data reuse.

Since only corresponding queries to specific entry vectors are used for distance calculations, non-active queries (those processed in other blocks) are skipped (line 9-11). This selective computation reduces unnecessary memory loads and calculations.
For matching queries, the algorithm computes partial distances between the query and the entry vectors (line 14). Finally, these partial distances are accumulated into the final distances array (line 15), completing the computation for the current block.
The result is a highly optimized distance calculation, Like GEMM, this algorithm exploits multiple levels of parallelism, it divides the computation into tiles and performs distance computations on these smaller tiles.

Note that this tiled computational pattern, cluster-centric rather than query-centric, is specifically tailored for entry point selection.
We assigns each cluster to a GPU block, which would be inefficient for traditional IVF workloads due to two factors.
First, the large number of IVF buckets (typically thousands) would cause significant read amplification when processed by GPU blocks (line 3).
Second, the high variance in IVF bucket sizes would lead to load imbalance, as the entire kernel must wait for the largest bucket to complete processing (line 16).

\textbf{Complexity analysis.}
As listed in Table~\ref{tab:complexity}, distance calculations of graph traversal has time complexity of $O(m n d)$.
For FES with $r$ clusters, the time complexity is $O(\frac{m n d}{r})$ as $n$ vectors are distributed to $r$ cells, where each query only performs $\frac{n}{r}$ computations.
When using 1 block, the FES reverts to brute force; and when there is only 1 query to process, the FES has the same complexity of graph traversal.
We set $r=32$ to match the GPU's warp size while obtaining a higher computational density.

%% file: resource/5-algorithm.tex
\begin{algorithm}[t]
    \caption{Tiled FES GPU Kernel}
    \newcommand{\PARFOR}{\STATE \textbf{parallel for}~}
    \newcommand{\ENDPARFOR}{\STATE \textbf{end parallel for}~}
    \begin{algorithmic}[1]
        \STATE {\bfseries Input:} queries $Q[m][d]$, entry vectors $EV[r][n/r][d]$
        \STATE {\bfseries Output:} distances $D[m][n]$
        \PARFOR{$block \gets$ loop $r$ GPU blocks}

        \FOR{$j \gets$ loop $n/r$ entry vectors, stride=32}
        \FOR{$k \gets$ loop vector dimension $d$, stride=32}
        \STATE $rhs \gets EV[block][j:j+32][k:k+32]$
        \PARFOR{$i \gets$ loop $m$ queries}
        \IF{$Q[i]$ not closest to $block$}
        \STATE $continue$ \COMMENT{Skip non-active query}
        \ENDIF
        \STATE $lhs[i\%32] \gets Q[i][k:k+32]$
        \STATE $synchronize()$ \COMMENT{Memory barrier}
        \STATE $D_{partial} \gets euclidean(lhs, rhs)$
        \STATE $D[i][j] \gets D[i][j] + D_{partial}$
        \ENDPARFOR
        \STATE $synchronize()$ \COMMENT{Wait all queries to finish}
        \ENDFOR
        \ENDFOR
        \ENDPARFOR
    \end{algorithmic}
    \label{alg:fast-routing}
\end{algorithm}

%% file: section/6-evaluation.tex
\section{Evaluation}\label{sec:evaluation}

\input{resource/6-eval-overall.tex}

\subsection{Experimental Setup}\label{subsec:eval-setup}

The experiments were conducted on cloud virtual machines equipped with Intel Xeon Platinum 8369B CPUs and an NVIDIA A10 GPU (24 GB). We allocated 32 vCPUs (equivalent to 16 OpenMP threads) for the evaluation.
Single-precision float32 is used for distance calculations on both CPU and GPU, and SIMD instructions up to AVX2 are utilized, following the performance tuning guidelines from the open source community~\footnote{https://github.com/facebookresearch/faiss/wiki/How-to-make-Faiss-run-faster}.

\textbf{Datasets.}
We use four 100-million-scale datasets listed in Table~\ref{tab:dataset}, DEEP~\cite{babenko2016deep1b}, T2I~\cite{yandex2021text2img}, WIKI, and LAION~\cite{schuhmann2022laion}, representing both moderate and high-dimensional data that exceed typical GPU memory capacities.
For the WIKI dataset, we constructed it by sampling 100 million paragraphs from English Wikipedia and generating embeddings using the BGE~\cite{xiao2023bge} model, which serves as the default embedding method in the LlamaIndex~\cite{liu2022llamaindex} framework.
We also employ smaller 1 million subsets of the same datasets (e.g., DEEP-1M) for analysis, as their original sizes can pose challenges.

\textbf{Baselines.}
We compare \name{} against the following state-of-the-art approaches:
\squishlist
\item \textit{HNSW}~\cite{malkov2018hnsw} is the industry-standard solution known for its robust performance. We evaluate the CPU implementation from FAISS~\cite{douze2024faiss}~1.8.0~\footnote{Recent updates to FAISS 1.8.0 have significantly enhanced its performance, making it a strong baseline for comparison: https://github.com/facebookresearch/faiss/pull/2841.}. To ensure a fair comparison, \name{} utilizes the same trained graph index as FAISS without altering the training algorithms.
\item \textit{RUMMY}~\cite{zhang2024rummy} is a recent IVF-based (coarse clustering) approach tackling similar challenges. It shares our focus on addressing GPU memory constraints in ANNS.
\item \textit{CAGRA}~\cite{ootomo2023cagra}, NVIDIA's ANNS algorithm, was evaluated as an additional baseline but encountered GPU memory boundaries when processing our datasets.
\squishend

\textbf{Metrics.}
We focus on recall as our primary metric, measured as:
$\text{recall}@k = \frac{| \text{retrieved}_k \cap \text{groundtruth}_k |}{k}$.
In particular, we evaluate the effectiveness of \name{} through recall-throughput curves, showing how accuracy trades off with computational performance.

\input{resource/6-eval-table.tex}

\subsection{Overall Performance}\label{subsec:eval-overall}

\name{} demonstrates significant performance improvements over the baselines across all scenarios.

\textbf{Overall.}
Figure~\ref{fig:eval-overall} illustrates the throughput improvements by \name{} compared to the \textit{HNSW-CPU} on different large scale datasets.
For the 96-dimensional DEEP dataset, our method achieves a $3.9 \times$ speedup compared to the baseline. Performance gains are even more significant for other datasets, showing $5.1 - 5.4 \times$ speedups, with the benefits becoming more pronounced as vector dimensions increase.
\textit{RUMMY} is slower than \textit{HNSW-CPU} in all scenarios because the IVF approach (which \textit{RUMMY} utilizes) inherently necessitates more distance calculations than HNSW.
Similar benefits for top-$100$ searches have also been observed (omit due to page limits).
Notably, T2I is a known difficult evaluation~\cite{kiela2022bigann}, our method delivers substantial speedups, despite not being specifically optimized for this dataset.

\textbf{Cost-effectiveness.}
Despite the significantly higher cost of the GPU-based platform (2.81 USD/hour) compared to the CPU-only solution (1.69 USD/hour), we achieve notable cost-effectiveness: $2.3 \times$ for DEEP, $3.0-3.2 \times$ for T2I, WIKI, and LAION in throughput per dollar, as illustrated in Table~\ref{tab:throughput-price}.
This suggests that \name{} excels with higher-dimensional ANNS.
Given the difficulties in improving CPU capabilities, \name{} offers a practical solution for enhancing ANNS processing capacity.

\input{resource/6-eval-scaling.tex}

\textbf{CAGRA-UVM.}
Figure~\ref{fig:eval-uvm} illustrates the comparison over \textit{CAGRA}.
\textit{CAGRA} does not scale beyond GPU memory limits (i.e., our datasets), we switch its memory allocations to \textit{cudaMallocManaged}~\cite{nvidia2020rmm}. This enables unified virtual memory (UVM), resulting in memory copying between the CPU and GPU when out-of-bounds access happens.
For datasets that fit within GPU memory, CAGRA attains up to $32 \times$ speedup over the baseline, while \name{} performs slightly worse because our kernel is designed to process subgraphs, resulting in additional overhead.
When the dataset exceeds GPU memory, UVM usage leads to inefficient CPU-GPU data transfers, causing slowdowns on \textit{CAGRA} that falls behind CPU-only approach.
In contrast, \name{} demonstrates scalability in handling growing datasets.

\textbf{Minimum GPU memory requirement.}
We evaluate the scalability of \name{} by determining the minimum GPU memory needed to achieve speedup.
We allocate up to 81\% of the available GPU memory (Table~\ref{tab:dataset}) for the graph index and vectors, reserving the remaining for dynamic allocations during query processing (e.g., query vectors and candidate sets).
Experiments with the LAION dataset demonstrates \name{}'s exceptional scalability.
With 19.4 GB GPU memory, we achieve a $4.8 \times$ throughput speedup while processing a dataset $14.9 \times$ larger than the the memory allocated.
As we decrease the GPU memory to 9.7 GB (dataset $29.7 \times$ larger), the system still maintains a $2.6 \times$ speedup.
As our approach is complementary to quantization techniques, future integration could enable even greater scalability.

\subsection{Component Analysis}\label{subsec:eval-components}

\begin{table}[t]
    \small
    \centering
    \setlength{\tabcolsep}{7.5mm}
    \caption{Ablation study on LAION dataset.}
    \begin{tabular}{ll}
        \toprule
        \textbf{Scheme}                         & \textbf{Throughput} \\
        \midrule
        \name{}                                 & 11,285              \\
        --~~CPU-GPU pipelining                  & 9,436 (-87.9\%)     \\
        --~~Fast entry selection                & 8,756 (-32.3\%)     \\
        --~~\textcircled{2} Residual refinement & 8,479 (-13.2\%)     \\
        --~~\textcircled{1} GPU piloting        & 2671 (-276\%)       \\
        FAISS                                   & 2,103               \\
        \bottomrule
    \end{tabular}
    \label{tab:ablation}
\end{table}

\textbf{Ablation study.}
We conducted an ablation study on the LAION dataset to evaluate CPU-GPU pipelining, fast entry selection, and the multi-stage processing.
As shown in Table~\ref{tab:ablation}, we sequentially removed each component and found that all components contribute positively to system performance.
Among these, the GPU piloting stage demonstrates the most substantial impact with a 276\% improvement.
Even when all components were removed, leaving only a naive CPU-only greedy search implementation, \name{} maintains a $1.27 \times$ speedup due to our hand-optimized AVX2 implementation.

Additional analyses can be found in Appendix~\S\ref{sec:appendix}.

%% file: resource/6-eval-overall.tex
\begin{figure*}[t]
    \centering
    \scriptsize
    \begin{subfigure}[t]{0.23\textwidth}
        \begin{tikzpicture}
            \begin{axis}[
                    ymode=log,
                    xlabel=Recall@10,
                    ylabel=Throughput (K),
                    xtick distance=0.05,
                    xmin=0.75, xmax=0.99,
                    ymin=0.50, ymax=150.50,
                    scale only axis=true,
                    width=0.80\textwidth,
                    height=0.50\textwidth,
                    legend cell align={center},
                    legend style={
                            anchor=south west, at={(1.60, 1.05)}, legend columns=4
                        }
                ]
                \addplot+ [color=brown!90, mark=+] coordinates {
                        (0.796, 1.41) (0.872, 1.25) (0.926, 0.97)
                        (0.943, 0.84) (0.957, 0.632)
                    };
                \addplot+ [smooth, color=orange!90, mark=x] coordinates {
                        (0.787, 16.3) (0.839, 14.76) (0.885, 12.88) (0.913, 10.33)
                        (0.931, 8.23) (0.942, 6.92) (0.953, 5.79) (0.959, 4.91)
                    };
                \addplot+ [smooth, color=cyan!90, mark=triangle] coordinates {
                        (0.831, 76.81) (0.860, 69.89) (0.884, 55.68) (0.905, 41.12)
                        (0.919, 32.70) (0.936, 22.24) (0.948, 17.21) (0.959, 11.93)
                    };
                \legend{RUMMY, HNSW-CPU, \name{}}
            \end{axis}
        \end{tikzpicture}
        \caption{DEEP}
    \end{subfigure}
    \hspace{0.01\textwidth}
    \begin{subfigure}[t]{0.23\textwidth}
        \begin{tikzpicture}
            \begin{axis}[
                    ymode=log,
                    xlabel=Recall@10,
                    ylabel=Throughput (K),
                    xtick distance=0.10,
                    xmin=0.50, xmax=0.99,
                    ymin=0.20, ymax=150.50,
                    scale only axis=true,
                    width=0.80\textwidth,
                    height=0.50\textwidth
                ]
                \addplot+ [color=brown!90, mark=+] coordinates {
                        (0.547, 1.23) (0.679, 0.96) (0.783, 0.67)
                        (0.828, 0.45) (0.853, 0.34)
                    };
                \addplot+ [smooth, color=orange!90, mark=x] coordinates {
                        (0.591, 7.96) (0.650, 7.41) (0.719, 6.54) (0.773, 5.25) (0.816, 3.99)
                        (0.851, 2.86) (0.876, 2.08) (0.889, 1.74) (0.895, 1.46) (0.901, 1.29)
                    };
                \addplot+ [smooth, color=cyan!90, mark=triangle] coordinates {
                        (0.672, 34.61) (0.724, 28.65) (0.757, 23.65) (0.796, 17.64) (0.820, 14.25)
                        (0.852, 10.27) (0.881, 7.63) (0.903, 5.77) (0.911, 5.17)
                    };
            \end{axis}
        \end{tikzpicture}
        \caption{T2I}
    \end{subfigure}
    \hspace{0.01\textwidth}
    \begin{subfigure}[t]{0.23\textwidth}
        \begin{tikzpicture}
            \begin{axis}[
                    ymode=log,
                    xlabel=Recall@10,
                    ylabel=Throughput (K),
                    xtick distance=0.05,
                    xmin=0.75, xmax=0.99,
                    ymin=0.20, ymax=150.50,
                    scale only axis=true,
                    width=0.80\textwidth,
                    height=0.50\textwidth
                ]
                \addplot+ [color=brown!90, mark=+] coordinates {
                        (0.823, 1.62) (0.884, 1.31) (0.916, 1.07)
                        (0.929, 0.89) (0.939, 0.67)
                    };
                \addplot+ [smooth, color=orange!90, mark=x] coordinates {
                        (0.788, 12.40) (0.828, 9.26) (0.861, 6.82) (0.890, 4.80)
                        (0.907, 3.83) (0.921, 3.10) (0.933, 2.47) (0.942, 1.91) (0.950, 1.45)
                    };
                \addplot+ [smooth, color=cyan!90, mark=triangle] coordinates {
                        (0.825, 43.32) (0.855, 34.32) (0.886, 25.32) (0.908, 19.11) (0.919, 16.01)
                        (0.927, 13.31) (0.936, 11.24) (0.943, 9.35) (0.948, 8.10) (0.953, 6.49)
                    };
            \end{axis}
        \end{tikzpicture}
        \caption{WIKI}
    \end{subfigure}
    \hspace{0.01\textwidth}
    \begin{subfigure}[t]{0.23\textwidth}
        \begin{tikzpicture}
            \begin{axis}[
                    ymode=log,
                    xlabel=Recall@10,
                    ylabel=Throughput (K),
                    xtick distance=0.10,
                    xmin=0.50, xmax=0.99,
                    ymin=0.20, ymax=150.50,
                    scale only axis=true,
                    width=0.80\textwidth,
                    height=0.50\textwidth
                ]
                \addplot+ [color=brown!90, mark=+] coordinates {
                        (0.641, 1.92) (0.776, 1.42) (0.820, 1.27) (0.864, 1.11) (0.906, 0.67)
                    };
                \addplot+ [smooth, color=orange!90, mark=x] coordinates {
                        (0.632, 14.45) (0.704, 10.12) (0.754, 7.72) (0.792, 5.95) (0.836, 4.13)
                        (0.861, 3.24) (0.879, 2.65) (0.892, 2.34) (0.903, 1.96) (0.912, 1.72)
                        (0.921, 1.52) (0.927, 1.24)
                    };
                \addplot+ [smooth, color=cyan!90, mark=triangle] coordinates {
                        (0.674, 50.18) (0.719, 41.34) (0.765, 32.65) (0.813, 23.41) (0.855, 17.34)
                        (0.887, 13.01) (0.913, 9.56) (0.925, 7.56) (0.932, 6.42) (0.941, 5.42)
                    };
            \end{axis}
        \end{tikzpicture}
        \caption{LAION}
    \end{subfigure}
    \caption{Recall-Throughput curves on 100 million datasets.}
    \label{fig:eval-overall}
\end{figure*}
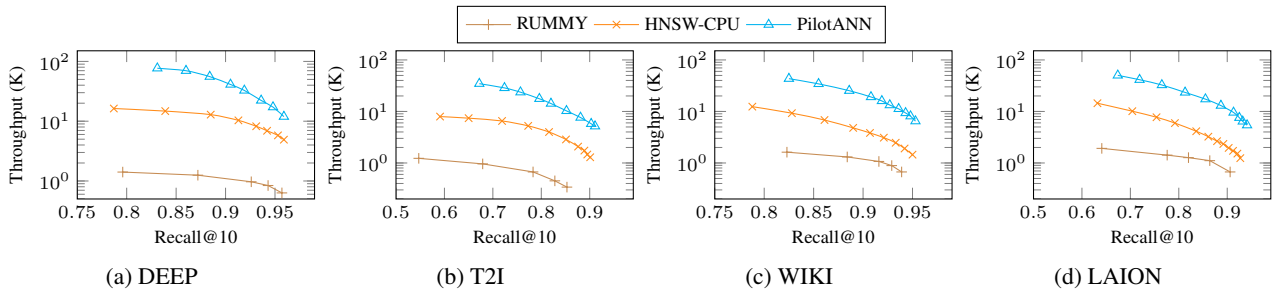

%% file: resource/6-eval-table.tex
\begin{table}[t]
    \small
    \centering
    \setlength{\tabcolsep}{1.5mm}
    \caption{Evaluation datasets.}
    \begin{tabular}{c|ccccc}
        \toprule
        \textbf{Dataset} & \textbf{Index size} & \textbf{Smpl. / SVD ratio} & \textbf{GPU mem.}       \\
        \midrule
        DEEP             & 59.6 GB             & 0.33, 1.0                  & 19.7 GB ($3.0 \times$)  \\
        T2I              & 98.3 GB             & 0.25, 0.64                 & 17.9 GB ($5.5 \times$)  \\
        WIKI             & 166.9 GB            & 0.25, 0.33                 & 17.9 GB ($9.3 \times$)  \\
        LAION            & 288.2 GB            & 0.25, 0.21                 & 19.4 GB ($14.9 \times$) \\
        \bottomrule
    \end{tabular}
    \label{tab:dataset}
\end{table}

\begin{table}[t]
    \small
    \centering
    \setlength{\tabcolsep}{1.5mm}
    \caption{Throughput and cost-effectiveness.}
    \begin{tabular}{c|cccc}
        \toprule
        \textbf{Dataset} & \textbf{Recall@10} & \textbf{FAISS} & \textbf{\name{}} & \textbf{Speedup. per \$} \\
        \midrule 
        \multirow{3}{*}{DEEP}
                         & 0.85               & 14,310         & 81,350           & $3.4 \times$             \\
                         & 0.90               & 11,514         & 44,586           & $2.3 \times$             \\
                         & 0.95               & 6,098          & 16,248           & $1.6 \times$             \\
        \midrule
        \multirow{3}{*}{T2I}
                         & 0.80               & 4,333          & 17,075           & $2.4 \times$             \\
                         & 0.85               & 2,869          & 10,519           & $2.2 \times$             \\
                         & 0.90               & 1,318          & 6,642            & $3.0 \times$             \\
        \midrule
        \multirow{3}{*}{WIKI}
                         & 0.85               & 7,633          & 35,821           & $2.8 \times$             \\
                         & 0.90               & 4,229          & 17,796           & $2.5 \times$             \\
                         & 0.95               & 1.448          & 7,456            & $3.1 \times$             \\
        \midrule
        \multirow{3}{*}{LAION}
                         & 0.80               & 5,619          & 25,912           & $2.8 \times$             \\
                         & 0.85               & 3,632          & 18,063           & $3.0 \times$             \\
                         & 0.90               & 2,103          & 11,285           & $3.2 \times$             \\
        \bottomrule
    \end{tabular}
    \label{tab:throughput-price}
\end{table}

%% file: resource/6-eval-scaling.tex
\begin{figure}[t]
    \centering
    \scriptsize
    \begin{minipage}[b]{0.22\textwidth}
        \begin{tikzpicture}
            \begin{axis}[
                    ybar, ymode=log,
                    width=1.20\textwidth,
                    height=0.75\textwidth,
                    enlarge x limits=0.30,
                    ylabel={Throughput (K)},
                    symbolic x coords={1M, 10M, 100M},
                    xtick=data, bar width=6pt, ymin=1.0, ymax=1000.0,
                    legend style={
                            anchor=north, at={(0.5, 1.55)},
                            legend columns=3, legend cell align={left}
                        },
                    legend image code/.code={
                            \draw[#1] (0cm,-0.1cm) rectangle (0.2cm,0.1cm);
                        }
                ]
                \addplot+ [color=black, fill=brown!60] coordinates {
                        (1M, 421.2) (10M, 1.2) (100M, 1.2)
                    };
                \addplot+ [color=black, fill=orange!60] coordinates {
                        (1M, 13.0) (10M, 7.9) (100M, 2.1)
                    };
                \addplot+ [color=black, fill=cyan!60] coordinates {
                        (1M, 285.7) (10M, 37.9) (100M, 11.3)
                    };
                \legend{CAGRA, CPU, \name{}}
            \end{axis}
        \end{tikzpicture}
        \caption{CAGRA on LAION dataset with UVM enabled.}
        \label{fig:eval-uvm}
    \end{minipage}
    \hfill
    \begin{minipage}[b]{0.22\textwidth}
        \begin{tikzpicture}
            \begin{axis}[
                    width=1.20\textwidth,
                    height=0.75\textwidth,
                    ylabel=Throughput (K), xlabel=GB,
                    legend style={
                            anchor=north, at={(0.5, 1.55)},
                            legend columns=2, legend cell align={left}
                        },
                ]
                \addplot+ [color=orange, mark=none, line width=0.5pt] coordinates {
                        (6.9, 4.2) (19.4, 4.2)
                    };
                \addplot+ [color=cyan, mark=diamond, line width=0.5pt] coordinates {
                        (6.9, 2.7) (8.3, 2.7) (9.7, 5.5) (13.9, 9.1) (16.6, 10.4) (19.4, 11.3)
                    };
                \legend{$2 \times$, \name{}}
            \end{axis}
        \end{tikzpicture}
        \caption{Minimum required GPU memory on LAION.}
    \end{minipage}
\end{figure}

%% file: section/7-related.tex
\section{Related Work}\label{sec:related}

\subsection{Large-scale ANNS}

Prior ANNS research focused primarily on index structure and data encoding optimizations, the Inverted Multi-Index (IMI)~\cite{babenko2015imi} enhanced space partitioning through multi-codebook quantization, while PQFastScan~\cite{andre2016fastscan} improved performance via SIMD and cache-aware optimizations.
FAISS~\cite{douze2024faiss}, a widely-adopted ANNS library, scales effectively to RAM capacity but has limited GPU support. DiskANN~\cite{jayaram2019diskann} and SPANN~\cite{chen2021spann} introduced novel disk-based indexing for billion-scale datasets, addressing different but related memory hierarchy challenges compared to our work.
\name{} is orthogonal to quantization and disk-based approaches, suggesting potential future integration of both approaches.

\subsection{GPU-accelerated ANNS}

Several approaches have leveraged GPUs for ANNS acceleration.
RUMMY~\cite{zhang2024rummy} implemented CPU-GPU pipelining for IVF-based search, though this strategy is suboptimal for graph-based indexes.
SONG~\cite{zhao2020song} and CAGRA~\cite{ootomo2023cagra} achieved significant speedups through GPU parallelization but their methods are constrained by GPU memory capacity.
BANG~\cite{karthik2024bang} handled billion-scale datasets using hybrid CPU-GPU processing but lacking CPU baseline comparisons~\footnote{Our claims may be outdated, as BANG is a work in progress.}.
In contrast, \name{} presents a memory-bounded GPU acceleration framework that effectively utilizes commodity GPUs while overcoming their memory constraints.
In addition, \name{} also maintains full compatibility with existing CPU systems and achieves significant speedups without compromising search accuracy or requiring high-end GPU hardware.

%% file: section/8-conclusion.tex
\section{Conclusion}

This work introduces a novel graph-based ANNS system that effectively utilizes both CPU and GPU for emerging ANNS workloads.
By decomposing top-$k$ search into a multi-stage CPU-GPU pipeline and employing efficient entry selection, our system achieves significant performance improvements over existing CPU-only approaches.

%% file: section/9-impact.tex
\section*{Impact Statement}

The effectiveness and efficiency of our proposed \name{} system democratizes high-performance nearest neighbor search by achieving competitive performance with just a single commodity GPU.
Our design significantly reduces computational overhead, making advanced search capabilities accessible to researchers and organizations with limited computing resources.
Unlike existing solutions that require expensive high-end GPUs, our approach enables efficient ANNS deployment on common hardware setups while maintaining search accuracy.
This work contributes to sustainable AI infrastructure development and empowers broader adoption of ANNS technology across diverse applications, especially for recent generative AIs.

%% file: section/10-appendix.tex
\section{Appendix}~\label{sec:appendix}

\subsection{Component Analysis}

\input{resource/10-eval-component.tex}

We evaluate the effectiveness of our proposed techniques.

\textbf{GPU piloting analysis.}
To handle large graph index within limited GPU memory, \name{} employs subgraph sampling and dimensionality reduction (\S\ref{sec:piloting}).
We evaluate how graph sampling and dimension reduction impact system performance on the LAION-1M dataset.
When using the full graph and original vectors (sampling and SVD ratio = 1.0) in GPU-only search, we observe a maximum speedup of $23.1 \times$ compared to the CPU baseline.
Using dimension reduction only (ratio = 0.25) achieves a $7.5 \times$ speedup, while using sampling only (ratio = 0.25) yields a $4.9 \times$ speedup.
This asymmetric impact suggests that our method tolerates aggressive dimension reduction while requiring a higher sampling ratio to maintain search quality.

\textbf{FES analysis.}
We evaluate the effectiveness of FES (\S\ref{sec:routing}) on the LAION-1M dataset using top-1000 recall as our metric, with the first 2-hop traversal of HNSW as the baseline, both evaluated on the GPU.
FES achieves 2017.0K QPS to reach a top-1000 recall of 0.001. This throughput represents a $16.2 \times$ speedup compared to the graph traversal baseline, which only obtains 124.7K QPS.
These results demonstrate that FES significantly improves both entry point quality and computational efficiency, making it a crucial component of \name{}.

\subsection{Sensitivity Analysis}

\input{resource/10-eval-sensitivity.tex}

To explore the sensitivity, we conducted more studies on the 1M datasets. The analysis highlights key observations regarding consistent benefits across graph construction methods and hardware settings.

\textbf{Graph construction.}
\name{} demonstrates orthogonality to graph construction methods, we compare the speedup using both HNSW and NSG graph construction methods.
While both combinations showed significant improvements over their baseline counterparts, \name{}+HNSW achieved higher speedups of $2.4\times$ and $4.8\times$ on DEEP-1M and LAION-1M respectively, compared to $2.2\times$ and $4.1\times$ of \name{}+NSG.
This is because NSG brings better search efficiency to the baseline, as the search quality and performance improve, different methods converge, leaving less room for relative improvements.

\textbf{Hardware independence.}
To validate the broad applicability of \name{}, we evaluated the system on Intel Xeon Platinum 8163 CPUs paired with an older NVIDIA T4 GPU, achieving $1.9\times$ and $4.5\times$ speedups on DEEP-1M and LAION-1M respectively, compared to $2.4\times$ and $4.8\times$ on our primary A10 test platform.
While the speedups are moderately lower on the T4 due to its reduced GPU compute capacity and PCIe bandwidth, the consistent benefits across both platforms confirm \name{}'s adaptability to different hardware architectures.

\subsection{Implementation}

\name{} is primarily built from scratch, comprising about 3.5K lines of Python, 1K lines of C++, and 1K lines of CUDA code.
This implementation was necessary due to the challenges in reusing existing ANNS libraries for CPU-GPU collaborating.
Notably, the system is implemented as an extension of LibTorch, which allows us to leverage the powerful tensor library for efficient CPU-GPU data management, facilitating seamless integration with deep learning models.
The CPU kernel matches the performance of the latest FAISS implementation, while the GPU kernel has been tailored to handle subgraph traversal, demonstrating improved capabilities over existing implementations.

%% file: resource/10-eval-component.tex
\begin{figure}[t]
    \centering
    \scriptsize
    \begin{minipage}{0.23\textwidth}
        \begin{tikzpicture}
            \pgfplotsset{
                colormap={cyanmap}{
                        rgb=(0.975, 0.975, 0.975)
                        rgb=(0.000, 0.718, 0.922)
                    }
            }
            \begin{axis}[
                    colorbar,
                    colormap name=cyanmap,
                    ylabel={SVD ratio},
                    xlabel={Sampling ratio},
                    width=0.95\textwidth,
                    height=0.95\textwidth,
                    xtick distance=0.25,
                    ytick distance=0.25,
                    xmin=0.125, xmax=1.125,
                    ymin=0.125, ymax=1.125,
                    colorbar style={width=2.5mm},
                    point meta min=0
                ],
                \addplot[
                    matrix plot*,
                    point meta=explicit,
                    mesh/cols=4
                ] table[meta=z] {
                        x y z
                        0.25 0.25 3.4
                        0.50 0.25 4.1
                        0.75 0.25 4.7
                        1.00 0.25 7.5
                        0.25 0.50 3.8
                        0.50 0.50 4.8
                        0.75 0.50 6.1
                        1.00 0.50 9.8
                        0.25 0.75 4.4
                        0.50 0.75 5.5
                        0.75 0.75 6.9
                        1.00 0.75 13.2
                        0.25 1.00 4.9
                        0.50 1.00 5.7
                        0.75 1.00 7.3
                        1.00 1.00 23.1
                    };
            \end{axis}
        \end{tikzpicture}
        \caption{Speedup on different GPU piloting configuration.}
        \label{fig:eval-gpu-heatmap}
    \end{minipage}
    \hfill
    \begin{minipage}{0.23\textwidth}
        \begin{tikzpicture}
            \begin{axis}[
                    ymode=log,
                    ylabel=Throughput (K),
                    xlabel=Recall@1000,
                    xtick distance=0.001,
                    ymin=10, ymax=10000.0,
                    scale only axis=true,
                    width=0.75\textwidth,
                    height=0.45\textwidth,
                    legend cell align={center},
                    legend style={
                            anchor=north, at={(0.5, 1.37)}, legend columns=2
                        }
                ]
                \addplot+ [smooth, color=orange, mark=x, line width=0.5pt] coordinates {
                        (0.0005, 169.8) (0.0010, 124.7) (0.0014, 91.9) (0.0017, 67.6) (0.0021, 32.1)
                    };
                \addplot+ [smooth, color=cyan, mark=triangle, line width=0.5pt] coordinates {
                        (0.0011, 2017.0) (0.0021, 1505.9) (0.0029, 966.0) (0.0036, 328.2) (0.0041, 90.2)
                    };
                \legend{HNSW-2-hop, FES}
            \end{axis}
        \end{tikzpicture}
        \caption{FES benefit.}
        \label{fig:eval-fes-benefit}
    \end{minipage}
\end{figure}

%% file: resource/10-eval-sensitivity.tex
\begin{figure}[t]
    \centering
    \scriptsize
    \begin{minipage}{0.23\textwidth}
        \centering
        \begin{tikzpicture}
            \begin{axis}[
                    ybar,
                    width=1.10\textwidth,
                    height=0.80\textwidth,
                    enlarge x limits=0.50,
                    ylabel={Speedup}, ymajorgrids,
                    symbolic x coords={DEEP-1M, LAION-1M},
                    xtick=data, bar width=6pt, 
                    legend style={
                            anchor=north, at={(0.5, 1.60)},
                            legend columns=2, legend cell align={left}
                        },
                    legend image code/.code={
                            \draw[#1, draw=none] (0cm,-0.1cm) rectangle (0.2cm,0.1cm);
                        }
                ]
                \addplot+ [
                    color=black, fill=orange!60
                ] coordinates {(DEEP-1M, 1.00) (LAION-1M, 1.00)};
                \addplot+ [
                    color=black, fill=cyan!60
                ] coordinates {(DEEP-1M, 2.37) (LAION-1M, 4.82)};
                \addplot+ [
                    color=black, fill=orange!60,
                    postaction={pattern=north west lines}
                ] coordinates {(DEEP-1M, 1.12) (LAION-1M, 1.33)};
                \addplot+[
                    color=black, fill=cyan!60,
                    postaction={pattern=north west lines}
                ] coordinates {(DEEP-1M, 2.16) (LAION-1M, 4.05)};
                \legend{HNSW, HNSW+\name{}, NSG, NSG+\name{}}
            \end{axis}
        \end{tikzpicture}
        \caption{HNSW v.s. NSG.}
        \label{fig:eval-nsg}
    \end{minipage}
    \hfill
    \begin{minipage}{0.23\textwidth}
        \centering
        \begin{tikzpicture}
            \begin{axis}[
                    ybar,
                    width=1.10\textwidth,
                    height=0.80\textwidth,
                    enlarge x limits=0.50,
                    ylabel={Speedup}, ymajorgrids,
                    symbolic x coords={DEEP-1M, LAION-1M},
                    xtick=data, bar width=6pt, 
                    legend style={
                            anchor=north, at={(0.5, 1.60)},
                            legend columns=2, legend cell align={left}
                        },
                    legend image code/.code={
                            \draw[#1, draw=none] (0cm,-0.1cm) rectangle (0.2cm,0.1cm);
                        }
                ]
                \addplot+ [
                    color=black, fill=orange!60,
                    postaction={pattern=north west lines}
                ] coordinates {(DEEP-1M, 0.92) (LAION-1M, 0.73)};
                \addplot+ [
                    color=black, fill=cyan!60,
                    postaction={pattern=north west lines}
                ] coordinates {(DEEP-1M, 1.73) (LAION-1M, 3.26)};
                \addplot+ [
                    color=black, fill=orange!60
                ] coordinates {(DEEP-1M, 1.00) (LAION-1M, 1.00)};
                \addplot+ [
                    color=black, fill=cyan!60
                ] coordinates {(DEEP-1M, 2.37) (LAION-1M, 4.82)};
                \legend{
                    24 vCPUs, 24 vCPUs + T4, 32 vCPUs, 32 vCPUs + A10
                }
            \end{axis}
        \end{tikzpicture}
        \caption{T4 v.s. A10.}
        \label{fig:eval-t4gpu}
    \end{minipage}
\end{figure}

%% file: main.bbl
\begin{thebibliography}{43}
\providecommand{\natexlab}[1]{#1}
\providecommand{\url}[1]{\texttt{#1}}
\expandafter\ifx\csname urlstyle\endcsname\relax
  \providecommand{\doi}[1]{doi: #1}\else
  \providecommand{\doi}{doi: \begingroup \urlstyle{rm}\Url}\fi

\bibitem[Andr{\'e} et~al.(2016)Andr{\'e}, Kermarrec, and
  Le~Scouarnec]{andre2016fastscan}
Andr{\'e}, F., Kermarrec, A.-M., and Le~Scouarnec, N.
\newblock Cache locality is not enough: High-performance nearest neighbor
  search with product quantization fast scan.
\newblock In \emph{42nd International Conference on Very Large Data Bases},
  volume~9, pp.\ ~12, 2016.

\bibitem[Babenko \& Lempitsky(2015)Babenko and Lempitsky]{babenko2015imi}
Babenko, A. and Lempitsky, V.~S.
\newblock The inverted multi-index.
\newblock \emph{{IEEE} Trans. Pattern Anal. Mach. Intell.}, 37\penalty0
  (6):\penalty0 1247--1260, 2015.
\newblock \doi{10.1109/TPAMI.2014.2361319}.
\newblock URL \url{https://doi.org/10.1109/TPAMI.2014.2361319}.

\bibitem[Babenko \& Lempitsky(2016)Babenko and Lempitsky]{babenko2016deep1b}
Babenko, A. and Lempitsky, V.~S.
\newblock Efficient indexing of billion-scale datasets of deep descriptors.
\newblock In \emph{2016 {IEEE} Conference on Computer Vision and Pattern
  Recognition, {CVPR} 2016, Las Vegas, NV, USA, June 27-30, 2016}, pp.\
  2055--2063. {IEEE} Computer Society, 2016.
\newblock \doi{10.1109/CVPR.2016.226}.
\newblock URL \url{https://doi.org/10.1109/CVPR.2016.226}.

\bibitem[Bang(2023)]{bang2023gptcache}
Bang, F.
\newblock Gptcache: An open-source semantic cache for llm applications enabling
  faster answers and cost savings.
\newblock In \emph{Proceedings of the 3rd Workshop for Natural Language
  Processing Open Source Software (NLP-OSS 2023)}, pp.\  212--218, 2023.

\bibitem[Blattmann et~al.(2022)Blattmann, Rombach, Oktay, M{\"u}ller, and
  Ommer]{blattmann2022diffusion}
Blattmann, A., Rombach, R., Oktay, K., M{\"u}ller, J., and Ommer, B.
\newblock Retrieval-augmented diffusion models.
\newblock \emph{Advances in Neural Information Processing Systems},
  35:\penalty0 15309--15324, 2022.

\bibitem[Chen et~al.(2021)Chen, Zhao, Wang, Li, Liu, Li, Yang, and
  Wang]{chen2021spann}
Chen, Q., Zhao, B., Wang, H., Li, M., Liu, C., Li, Z., Yang, M., and Wang, J.
\newblock {SPANN:} highly-efficient billion-scale approximate nearest
  neighborhood search.
\newblock In Ranzato, M., Beygelzimer, A., Dauphin, Y.~N., Liang, P., and
  Vaughan, J.~W. (eds.), \emph{Advances in Neural Information Processing
  Systems 34: Annual Conference on Neural Information Processing Systems 2021,
  NeurIPS 2021, December 6-14, 2021, virtual}, pp.\  5199--5212, 2021.
\newblock URL
  \url{https://proceedings.neurips.cc/paper/2021/hash/299dc35e747eb77177d9cea10a802da2-Abstract.html}.

\bibitem[Covington et~al.(2016)Covington, Adams, and
  Sargin]{covington2016youtube}
Covington, P., Adams, J., and Sargin, E.
\newblock Deep neural networks for youtube recommendations.
\newblock In \emph{Proceedings of the 10th ACM conference on recommender
  systems}, pp.\  191--198, 2016.

\bibitem[Devlin et~al.(2018)Devlin, Chang, Lee, and Toutanova]{devlin2018bert}
Devlin, J., Chang, M.-W., Lee, K., and Toutanova, K.
\newblock Bert: Pre-training of deep bidirectional transformers for language
  understanding.
\newblock \emph{arXiv preprint arXiv:1810.04805}, 2018.

\bibitem[Dong et~al.(2011)Dong, Charikar, and Li]{dong2011nndescent}
Dong, W., Charikar, M., and Li, K.
\newblock Efficient k-nearest neighbor graph construction for generic
  similarity measures.
\newblock In Srinivasan, S., Ramamritham, K., Kumar, A., Ravindra, M.~P.,
  Bertino, E., and Kumar, R. (eds.), \emph{Proceedings of the 20th
  International Conference on World Wide Web, {WWW} 2011, Hyderabad, India,
  March 28 - April 1, 2011}, pp.\  577--586. {ACM}, 2011.
\newblock \doi{10.1145/1963405.1963487}.
\newblock URL \url{https://doi.org/10.1145/1963405.1963487}.

\bibitem[Dosovitskiy et~al.(2021)Dosovitskiy, Beyer, Kolesnikov, Weissenborn,
  Zhai, Unterthiner, Dehghani, Minderer, Heigold, Gelly, Uszkoreit, and
  Houlsby]{dosovitskiy2021vit}
Dosovitskiy, A., Beyer, L., Kolesnikov, A., Weissenborn, D., Zhai, X.,
  Unterthiner, T., Dehghani, M., Minderer, M., Heigold, G., Gelly, S.,
  Uszkoreit, J., and Houlsby, N.
\newblock An image is worth 16x16 words: Transformers for image recognition at
  scale.
\newblock In \emph{9th International Conference on Learning Representations,
  {ICLR} 2021, Virtual Event, Austria, May 3-7, 2021}. OpenReview.net, 2021.
\newblock URL \url{https://openreview.net/forum?id=YicbFdNTTy}.

\bibitem[Douze et~al.(2018)Douze, Sablayrolles, and
  J{\'{e}}gou]{douze2018linkandcode}
Douze, M., Sablayrolles, A., and J{\'{e}}gou, H.
\newblock Link and code: Fast indexing with graphs and compact regression
  codes.
\newblock In \emph{2018 {IEEE} Conference on Computer Vision and Pattern
  Recognition, {CVPR} 2018, Salt Lake City, UT, USA, June 18-22, 2018}, pp.\
  3646--3654. Computer Vision Foundation / {IEEE} Computer Society, 2018.
\newblock \doi{10.1109/CVPR.2018.00384}.
\newblock URL
  \url{http://openaccess.thecvf.com/content\_cvpr\_2018/html/Douze\_Link\_and\_Code\_CVPR\_2018\_paper.html}.

\bibitem[Douze et~al.(2024)Douze, Guzhva, Deng, Johnson, Szilvasy,
  Mazar{\'{e}}, Lomeli, Hosseini, and J{\'{e}}gou]{douze2024faiss}
Douze, M., Guzhva, A., Deng, C., Johnson, J., Szilvasy, G., Mazar{\'{e}}, P.,
  Lomeli, M., Hosseini, L., and J{\'{e}}gou, H.
\newblock The faiss library.
\newblock \emph{CoRR}, abs/2401.08281, 2024.
\newblock \doi{10.48550/ARXIV.2401.08281}.
\newblock URL \url{https://doi.org/10.48550/arXiv.2401.08281}.

\bibitem[Fu et~al.(2019)Fu, Xiang, Wang, and Cai]{fu2019nsg}
Fu, C., Xiang, C., Wang, C., and Cai, D.
\newblock Fast approximate nearest neighbor search with the navigating
  spreading-out graph.
\newblock \emph{Proc. {VLDB} Endow.}, 12\penalty0 (5):\penalty0 461--474, 2019.
\newblock \doi{10.14778/3303753.3303754}.
\newblock URL \url{http://www.vldb.org/pvldb/vol12/p461-fu.pdf}.

\bibitem[Gionis et~al.(1999)Gionis, Indyk, and Motwani]{gionis1999lsh}
Gionis, A., Indyk, P., and Motwani, R.
\newblock Similarity search in high dimensions via hashing.
\newblock In Atkinson, M.~P., Orlowska, M.~E., Valduriez, P., Zdonik, S.~B.,
  and Brodie, M.~L. (eds.), \emph{VLDB'99, Proceedings of 25th International
  Conference on Very Large Data Bases, September 7-10, 1999, Edinburgh,
  Scotland, {UK}}, pp.\  518--529. Morgan Kaufmann, 1999.
\newblock URL \url{http://www.vldb.org/conf/1999/P49.pdf}.

\bibitem[Gordo et~al.(2016)Gordo, Almaz{\'a}n, Revaud, and
  Larlus]{gordo2016search}
Gordo, A., Almaz{\'a}n, J., Revaud, J., and Larlus, D.
\newblock Deep image retrieval: Learning global representations for image
  search.
\newblock In \emph{Computer Vision--ECCV 2016: 14th European Conference,
  Amsterdam, The Netherlands, October 11-14, 2016, Proceedings, Part VI 14},
  pp.\  241--257. Springer, 2016.

\bibitem[Hajebi et~al.(2011)Hajebi, Abbasi{-}Yadkori, Shahbazi, and
  Zhang]{hajebi2011knngraph}
Hajebi, K., Abbasi{-}Yadkori, Y., Shahbazi, H., and Zhang, H.
\newblock Fast approximate nearest-neighbor search with k-nearest neighbor
  graph.
\newblock In Walsh, T. (ed.), \emph{{IJCAI} 2011, Proceedings of the 22nd
  International Joint Conference on Artificial Intelligence, Barcelona,
  Catalonia, Spain, July 16-22, 2011}, pp.\  1312--1317. {IJCAI/AAAI}, 2011.
\newblock \doi{10.5591/978-1-57735-516-8/IJCAI11-222}.
\newblock URL \url{https://doi.org/10.5591/978-1-57735-516-8/IJCAI11-222}.

\bibitem[He et~al.(2017)He, Liao, Zhang, Nie, Hu, and Chua]{he2017cf}
He, X., Liao, L., Zhang, H., Nie, L., Hu, X., and Chua, T.-S.
\newblock Neural collaborative filtering.
\newblock In \emph{Proceedings of the 26th international conference on world
  wide web}, pp.\  173--182, 2017.

\bibitem[Jayaram~Subramanya et~al.(2019)Jayaram~Subramanya, Devvrit, Simhadri,
  Krishnawamy, and Kadekodi]{jayaram2019diskann}
Jayaram~Subramanya, S., Devvrit, F., Simhadri, H.~V., Krishnawamy, R., and
  Kadekodi, R.
\newblock Diskann: Fast accurate billion-point nearest neighbor search on a
  single node.
\newblock \emph{Advances in Neural Information Processing Systems}, 32, 2019.

\bibitem[J{\'{e}}gou et~al.(2011{\natexlab{a}})J{\'{e}}gou, Douze, and
  Schmid]{jegou2010ivfadc}
J{\'{e}}gou, H., Douze, M., and Schmid, C.
\newblock Product quantization for nearest neighbor search.
\newblock \emph{{IEEE} Trans. Pattern Anal. Mach. Intell.}, 33\penalty0
  (1):\penalty0 117--128, 2011{\natexlab{a}}.
\newblock \doi{10.1109/TPAMI.2010.57}.
\newblock URL \url{https://doi.org/10.1109/TPAMI.2010.57}.

\bibitem[J{\'{e}}gou et~al.(2011{\natexlab{b}})J{\'{e}}gou, Tavenard, Douze,
  and Amsaleg]{jegou2011ivfadcr}
J{\'{e}}gou, H., Tavenard, R., Douze, M., and Amsaleg, L.
\newblock Searching in one billion vectors: Re-rank with source coding.
\newblock In \emph{Proceedings of the {IEEE} International Conference on
  Acoustics, Speech, and Signal Processing, {ICASSP} 2011, May 22-27, 2011,
  Prague Congress Center, Prague, Czech Republic}, pp.\  861--864. {IEEE},
  2011{\natexlab{b}}.
\newblock \doi{10.1109/ICASSP.2011.5946540}.
\newblock URL \url{https://doi.org/10.1109/ICASSP.2011.5946540}.

\bibitem[Karthik et~al.(2024)Karthik, Khan, Singh, Simhadri, and
  Vedurada]{karthik2024bang}
Karthik, V., Khan, S., Singh, S., Simhadri, H.~V., and Vedurada, J.
\newblock Bang: Billion-scale approximate nearest neighbor search using a
  single gpu.
\newblock \emph{arXiv e-prints}, pp.\  arXiv--2401, 2024.

\bibitem[Kulis \& Grauman(2009)Kulis and Grauman]{kulis2009search}
Kulis, B. and Grauman, K.
\newblock Kernelized locality-sensitive hashing for scalable image search.
\newblock In \emph{2009 IEEE 12th international conference on computer vision},
  pp.\  2130--2137. IEEE, 2009.

\bibitem[Lewis et~al.(2020)Lewis, Perez, Piktus, Petroni, Karpukhin, Goyal,
  K{\"{u}}ttler, Lewis, Yih, Rockt{\"{a}}schel, Riedel, and
  Kiela]{lewis2020rag}
Lewis, P. S.~H., Perez, E., Piktus, A., Petroni, F., Karpukhin, V., Goyal, N.,
  K{\"{u}}ttler, H., Lewis, M., Yih, W., Rockt{\"{a}}schel, T., Riedel, S., and
  Kiela, D.
\newblock Retrieval-augmented generation for knowledge-intensive {NLP} tasks.
\newblock In Larochelle, H., Ranzato, M., Hadsell, R., Balcan, M., and Lin, H.
  (eds.), \emph{Advances in Neural Information Processing Systems 33: Annual
  Conference on Neural Information Processing Systems 2020, NeurIPS 2020,
  December 6-12, 2020, virtual}, 2020.
\newblock URL
  \url{https://proceedings.neurips.cc/paper/2020/hash/6b493230205f780e1bc26945df7481e5-Abstract.html}.

\bibitem[Liu(2022)]{liu2022llamaindex}
Liu, J.
\newblock {LlamaIndex}, 11 2022.
\newblock URL \url{https://github.com/jerryjliu/llama_index}.

\bibitem[Lu et~al.(2021)Lu, Kudo, Xiao, and Ishikawa]{lu2021hvs}
Lu, K., Kudo, M., Xiao, C., and Ishikawa, Y.
\newblock Hvs: hierarchical graph structure based on voronoi diagrams for
  solving approximate nearest neighbor search.
\newblock \emph{Proc. VLDB Endow.}, 15\penalty0 (2):\penalty0 246–258,
  October 2021.
\newblock ISSN 2150-8097.
\newblock \doi{10.14778/3489496.3489506}.
\newblock URL \url{https://doi.org/10.14778/3489496.3489506}.

\bibitem[Malkov et~al.(2014)Malkov, Ponomarenko, Logvinov, and
  Krylov]{malkov2014nsw}
Malkov, Y., Ponomarenko, A., Logvinov, A., and Krylov, V.
\newblock Approximate nearest neighbor algorithm based on navigable small world
  graphs.
\newblock \emph{Information Systems}, 45:\penalty0 61--68, 2014.

\bibitem[Malkov \& Yashunin(2020)Malkov and Yashunin]{malkov2018hnsw}
Malkov, Y.~A. and Yashunin, D.~A.
\newblock Efficient and robust approximate nearest neighbor search using
  hierarchical navigable small world graphs.
\newblock \emph{{IEEE} Trans. Pattern Anal. Mach. Intell.}, 42\penalty0
  (4):\penalty0 824--836, 2020.
\newblock \doi{10.1109/TPAMI.2018.2889473}.
\newblock URL \url{https://doi.org/10.1109/TPAMI.2018.2889473}.

\bibitem[NVIDIA(2020)]{nvidia2020rmm}
NVIDIA.
\newblock {F}ast, {F}lexible {A}llocation for {N}{V}{I}{D}{I}{A} {C}{U}{D}{A}
  with {R}{A}{P}{I}{D}{S} {M}emory {M}anager | {N}{V}{I}{D}{I}{A} {T}echnical
  {B}log --- developer.nvidia.com.
\newblock
  \url{https://developer.nvidia.com/blog/fast-flexible-allocation-for-cuda-with-rapids-memory-manager},
  2020.
\newblock [Accessed 15-10-2024].

\bibitem[Ootomo et~al.(2023)Ootomo, Naruse, Nolet, Wang, Feher, and
  Wang]{ootomo2023cagra}
Ootomo, H., Naruse, A., Nolet, C., Wang, R., Feher, T., and Wang, Y.
\newblock {CAGRA:} highly parallel graph construction and approximate nearest
  neighbor search for gpus.
\newblock \emph{CoRR}, abs/2308.15136, 2023.
\newblock \doi{10.48550/ARXIV.2308.15136}.
\newblock URL \url{https://doi.org/10.48550/arXiv.2308.15136}.

\bibitem[Peng et~al.(2023)Peng, Zhang, Li, Jin, and Ren]{peng2023speedann}
Peng, Z., Zhang, M., Li, K., Jin, R., and Ren, B.
\newblock iqan: Fast and accurate vector search with efficient intra-query
  parallelism on multi-core architectures.
\newblock In Dehnavi, M.~M., Kulkarni, M., and Krishnamoorthy, S. (eds.),
  \emph{Proceedings of the 28th {ACM} {SIGPLAN} Annual Symposium on Principles
  and Practice of Parallel Programming, PPoPP 2023, Montreal, QC, Canada, 25
  February 2023 - 1 March 2023}, pp.\  313--328. {ACM}, 2023.
\newblock \doi{10.1145/3572848.3577527}.
\newblock URL \url{https://doi.org/10.1145/3572848.3577527}.

\bibitem[Radford et~al.(2021)Radford, Kim, Hallacy, Ramesh, Goh, Agarwal,
  Sastry, Askell, Mishkin, Clark, Krueger, and Sutskever]{radford2021clip}
Radford, A., Kim, J.~W., Hallacy, C., Ramesh, A., Goh, G., Agarwal, S., Sastry,
  G., Askell, A., Mishkin, P., Clark, J., Krueger, G., and Sutskever, I.
\newblock Learning transferable visual models from natural language
  supervision.
\newblock In Meila, M. and Zhang, T. (eds.), \emph{Proceedings of the 38th
  International Conference on Machine Learning, {ICML} 2021, 18-24 July 2021,
  Virtual Event}, volume 139 of \emph{Proceedings of Machine Learning
  Research}, pp.\  8748--8763. {PMLR}, 2021.
\newblock URL \url{http://proceedings.mlr.press/v139/radford21a.html}.

\bibitem[Schuhmann et~al.(2022)Schuhmann, Beaumont, Vencu, Gordon, Wightman,
  Cherti, Coombes, Katta, Mullis, Wortsman, Schramowski, Kundurthy, Crowson,
  Schmidt, Kaczmarczyk, and Jitsev]{schuhmann2022laion}
Schuhmann, C., Beaumont, R., Vencu, R., Gordon, C., Wightman, R., Cherti, M.,
  Coombes, T., Katta, A., Mullis, C., Wortsman, M., Schramowski, P., Kundurthy,
  S., Crowson, K., Schmidt, L., Kaczmarczyk, R., and Jitsev, J.
\newblock {LAION-5B:} an open large-scale dataset for training next generation
  image-text models.
\newblock In Koyejo, S., Mohamed, S., Agarwal, A., Belgrave, D., Cho, K., and
  Oh, A. (eds.), \emph{Advances in Neural Information Processing Systems 35:
  Annual Conference on Neural Information Processing Systems 2022, NeurIPS
  2022, New Orleans, LA, USA, November 28 - December 9, 2022}, 2022.
\newblock URL
  \url{http://papers.nips.cc/paper\_files/paper/2022/hash/a1859debfb3b59d094f3504d5ebb6c25-Abstract-Datasets\_and\_Benchmarks.html}.

\bibitem[Sch{\"u}tze et~al.(2008)Sch{\"u}tze, Manning, and
  Raghavan]{schutze2008introduction}
Sch{\"u}tze, H., Manning, C.~D., and Raghavan, P.
\newblock \emph{Introduction to information retrieval}, volume~39.
\newblock Cambridge University Press Cambridge, 2008.

\bibitem[Silpa{-}Anan \& Hartley(2008)Silpa{-}Anan and Hartley]{anan2008kdtree}
Silpa{-}Anan, C. and Hartley, R.~I.
\newblock Optimised kd-trees for fast image descriptor matching.
\newblock In \emph{2008 {IEEE} Computer Society Conference on Computer Vision
  and Pattern Recognition {(CVPR} 2008), 24-26 June 2008, Anchorage, Alaska,
  {USA}}. {IEEE} Computer Society, 2008.
\newblock \doi{10.1109/CVPR.2008.4587638}.
\newblock URL \url{https://doi.org/10.1109/CVPR.2008.4587638}.

\bibitem[Simhadri et~al.(2022)Simhadri, Williams, Aum\"uller, Douze, Babenko,
  Baranchuk, Chen, Hosseini, Krishnaswamny, Srinivasa, Subramanya, and
  Wang]{kiela2022bigann}
Simhadri, H.~V., Williams, G., Aum\"uller, M., Douze, M., Babenko, A.,
  Baranchuk, D., Chen, Q., Hosseini, L., Krishnaswamny, R., Srinivasa, G.,
  Subramanya, S.~J., and Wang, J.
\newblock Results of the neurips{’}21 challenge on billion-scale approximate
  nearest neighbor search.
\newblock In Kiela, D., Ciccone, M., and Caputo, B. (eds.), \emph{Proceedings
  of the NeurIPS 2021 Competitions and Demonstrations Track}, volume 176 of
  \emph{Proceedings of Machine Learning Research}, pp.\  177--189. PMLR, 06--14
  Dec 2022.
\newblock URL \url{https://proceedings.mlr.press/v176/simhadri22a.html}.

\bibitem[Williams et~al.(2009)Williams, Waterman, and
  Patterson]{williams2009roofline}
Williams, S., Waterman, A., and Patterson, D.~A.
\newblock Roofline: an insightful visual performance model for multicore
  architectures.
\newblock \emph{Commun. {ACM}}, 52\penalty0 (4):\penalty0 65--76, 2009.
\newblock \doi{10.1145/1498765.1498785}.
\newblock URL \url{https://doi.org/10.1145/1498765.1498785}.

\bibitem[Xiao et~al.(2023)Xiao, Liu, Zhang, and Muennighoff]{xiao2023bge}
Xiao, S., Liu, Z., Zhang, P., and Muennighoff, N.
\newblock C-pack: Packaged resources to advance general chinese embedding,
  2023.

\bibitem[Yandex(2021)]{yandex2021text2img}
Yandex.
\newblock {Y}andex --- research.yandex.com.
\newblock
  \url{https://research.yandex.com/blog/benchmarks-for-billion-scale-similarity-search},
  2021.
\newblock [Accessed 15-10-2024].

\bibitem[Yang et~al.(2020)Yang, Yi, Zhiyuan~Cheng, Hong, Li, Xiaoming~Wang, Xu,
  and Chi]{yang2020twotower}
Yang, J., Yi, X., Zhiyuan~Cheng, D., Hong, L., Li, Y., Xiaoming~Wang, S., Xu,
  T., and Chi, E.~H.
\newblock Mixed negative sampling for learning two-tower neural networks in
  recommendations.
\newblock In \emph{Companion proceedings of the web conference 2020}, pp.\
  441--447, 2020.

\bibitem[Zhang et~al.(2024)Zhang, Liu, Huang, Liu, and Jin]{zhang2024rummy}
Zhang, Z., Liu, F., Huang, G., Liu, X., and Jin, X.
\newblock Fast vector query processing for large datasets beyond $\{$GPU$\}$
  memory with reordered pipelining.
\newblock In \emph{21st USENIX Symposium on Networked Systems Design and
  Implementation (NSDI 24)}, pp.\  23--40, 2024.

\bibitem[Zhao et~al.(2020)Zhao, Tan, and Li]{zhao2020song}
Zhao, W., Tan, S., and Li, P.
\newblock {SONG:} approximate nearest neighbor search on {GPU}.
\newblock In \emph{36th {IEEE} International Conference on Data Engineering,
  {ICDE} 2020, Dallas, TX, USA, April 20-24, 2020}, pp.\  1033--1044. {IEEE},
  2020.
\newblock \doi{10.1109/ICDE48307.2020.00094}.
\newblock URL \url{https://doi.org/10.1109/ICDE48307.2020.00094}.

\bibitem[Zhou et~al.(2012)Zhou, Lu, Li, and Tian]{zhou2012scalar}
Zhou, W., Lu, Y., Li, H., and Tian, Q.
\newblock Scalar quantization for large scale image search.
\newblock In \emph{Proceedings of the 20th ACM international conference on
  Multimedia}, pp.\  169--178, 2012.

\bibitem[Zhu(2022)]{zhu2022rtnn}
Zhu, Y.
\newblock {RTNN:} accelerating neighbor search using hardware ray tracing.
\newblock In Lee, J., Agrawal, K., and Spear, M.~F. (eds.), \emph{PPoPP '22:
  27th {ACM} {SIGPLAN} Symposium on Principles and Practice of Parallel
  Programming, Seoul, Republic of Korea, April 2 - 6, 2022}, pp.\  76--89.
  {ACM}, 2022.
\newblock \doi{10.1145/3503221.3508409}.
\newblock URL \url{https://doi.org/10.1145/3503221.3508409}.

\end{thebibliography}
